\documentclass[journal]{IEEEtran}

\usepackage{cite}
\usepackage{amsmath,amssymb,amsfonts}
\usepackage{amsthm}
\usepackage{graphicx}
\usepackage{booktabs}
\usepackage{multirow}
\usepackage{array}
\usepackage{url}
\usepackage{tabularx}
\usepackage{hyperref}

\graphicspath{{figures/}}

\begin{document}

\title{A Wireless World Model for AI-Native 6G Networks}

\author{
Ziqi~Chen,
Yi~Ren,
Yixuan~Huang,
Qi~Sun,
Nan~Li,
Yuhong~Huang,
Yifan~Li,
and Liang~Xia
\thanks{Ziqi Chen, Yi Ren, Yixuan Huang, Qi Sun, Nan Li, Yuhong Huang, Yifan Li, and Liang Xia are with China Mobile Research Institute, Beijing 100053, China.}
\thanks{Corresponding authors: Ziqi Chen, Nan Li, and Yuhong Huang (e-mail: chenziqiyjy@chinamobile.com; linan@chinamobile.com; huangyuhong@chinamobile.com).}
\thanks{Ziqi Chen, Yi Ren, Yixuan Huang, and Qi Sun contributed equally to this work.}
}

\maketitle

\begin{abstract}
Integrating AI into the physical layer is a cornerstone of 6G networks. However, current data-driven approaches struggle to generalize across dynamic environments because they lack an intrinsic understanding of electromagnetic wave propagation. We introduce the Wireless World Model (WWM), a multi-modal foundation framework predicting the spatiotemporal evolution of wireless channels by internalizing the physical relationship between 3D geometry and signal dynamics. Pre-trained on a massive ray-traced multi-modal dataset, WWM overcomes the data authenticity gap, further validated under real-world measurement data. Using a joint-embedding predictive architecture with a multi-modal mixture-of-experts Transformer, WWM fuses channel state information, 3D point clouds, and user trajectories into a unified representation. Across the five key downstream tasks supported by WWM, it achieves remarkable performance in seen environments, unseen generalization scenarios, and real-world measurements, consistently outperforming SOTA uni-modal foundation models and task-specific models. This paves the way for physics-aware 6G intelligence that adapts to the physical world.
\end{abstract}


\section{Introduction}

Sixth-generation (6G) wireless networks are envisioned as the core information infrastructure for the AI era, necessitating a quantum leap in performance and new capabilities like integrated AI and communication~\cite{6G1, 6G2}. Central to this evolution is the enhancement of spectral efficiency at the air interface—the most fundamental and historically challenging goal of every mobile generation. The progression from 2G to 5G achieved this primarily by increasing bandwidth and deploying large-scale Multiple Input and Multiple Output (MIMO) antenna systems, guided by Shannon's information theory~\cite{shannon}. However, this traditional scaling approach has reached a bottleneck. Further expanding to extremely large massive MIMO systems introduces prohibitive overhead for acquiring precise Channel State Information (CSI) and intractable computational complexity for precoding~\cite{MIMO1,MIMO2}. Furthermore, a persistent gap to the theoretical Shannon limit exists, widened by hardware non-idealities and complex network interference that conventional signal processing models cannot adequately address~\cite{MIMO3}. AI presents a transformative solution, leveraging its advanced capabilities in feature extraction and complex problem-solving to break these barriers~\cite{AI6G1}. This propels the vision of an AI-native 6G air interface, where AI is fundamentally integrated into core physical layer functions~\cite{AI6G2,AI6G3,AI6G4}. 
Initial efforts toward an AI-native air interface relied on task-specific models. However, this approach is fundamentally limited by poor generalization to dynamic environments, and a disjointed design that creates significant computational and management burdens at the base station (BS). In recent years, foundation models have demonstrated superior performance and changed the landscape of AI~\cite{moor2023foundation,abramson2024accurate,he2025generalized,binz2025foundation}. This has catalyzed a paradigm shift toward a more scalable wireless AI solution: the Wireless Foundation Model (WFM). A WFM is a large-scale model, pre-trained on diverse wireless data via self-supervision to improve generalization capability, and designed to adapt to a wide range of downstream tasks with minimal fine-tuning.

Early explorations into WFMs initially focused on adapting pre-trained Large Language Models (LLMs) to wireless domain through fine-tuning or prompt engineering\cite{WFM0,llm4cp,WFM1,WFM2,WFM3,WFM4,WFM5}. While this demonstrated the potential of large-scale AI model architectures, the approach often yielded physically inconsistent predictions, as the models lacked an intrinsic understanding of electromagnetic wave propagation. This critical flaw spurred a necessary shift toward the current generation of AI-native architectures, such as WiFo \cite{wifo1,wifo2} and WirelessGPT \cite{wirelessGPT}, which are pre-trained from the ground up on vast, domain-specific corpora. Despite this progress, the field remains constrained by two limitations. First is a data authenticity gap; most models are trained on statistical channel data that fails to capture real-world complexity, a problem addressable by integrating high-fidelity, physics-based data like ray tracing~\cite{RT}. Second, existing WFMs make limited use of explicit three-dimensional geometric context. Recent studies~\cite{jiao2025addressing,yazdnian2026multi,zheng2026muse} have incorporated multimodal environmental context into wireless foundation models, yet their environmental representations remain map or hight map-based rather than explicit 3D geometric representations such as point clouds, Gaussian splats, or voxel grids, which are commonly used to represent explicit 3D scene geometry in computer vision, robotics, and physical AI. With recent technological advances, offline 3D mapping, LiDAR sensing, and city-scale modeling have become increasingly prevalent, making such semi-static 3D geometric information more readily available and is expected to play an increasingly important role in wireless networks.


To address these challenges, we propose the Wireless World Model (WWM), marking a shift from data-driven pattern matching to environment-aware cognitive intelligence. Conceptually, a World Model learns an internal predictive representation of the environment, enabling a system to understand the world it operates in and predict future states in a manner consistent with physical laws~\cite{NEURIPS2018_2de5d166, lecun2022path, ren2025cosmos}. While a WFM excels at capturing statistical correlations within signal distributions, the WWM aims to learn the wireless channel pattern conditioned on  surrounding 3D spatial geometry. By learning the relationship between environmental semantics and channel characteristics, the WWM generates a latent representation that captures how physical objects and spatial movements influence channel behavior. By perceiving the environment as a structured physical entity, the WWM exhibits three distinct features: environmental grounding, which links signal variations to physical objects; predictive consistency, ensuring channel estimations align with electromagnetic theory; and multi-task versatility, supporting diverse downstream functions through a shared understanding of the environment.

In this work, we make three key contributions. First, we construct a massive hybrid multi-modal dataset of approximately 800 thousand samples by integrating high-fidelity Sionna RT ray tracing simulations~\cite{sionna} with real-world 6G prototype BS field trials, ensuring that the model learns from both physically consistent and realistic electromagnetic environments. Second, the model adopts a Joint Embedding Predictive Architecture (JEPA)~\cite{bardes2024revisiting, assran2025v}, which shifts the learning objective from signal reconstruction to prediction in a semantic feature space. This compels the model to learn the physics-informed relationship between scene geometry, user motion, and electromagnetic propagation. Third, a Multi-Modal Mixture of Experts (MMoE) structure~\cite{wang2023image} is employed, enabling efficient fusion of multi-modal data including channel signal, point cloud and trajectory data within a unified Transformer backbone. This architecture establishes WWM as a shared pretrained representation backbone. The pretrained backbone is reused across downstream RAN applications and paired with lightweight task-specific heads, supporting tasks including channel prediction, CSI compression and feedback, beam prediction, user localization. As the first model to systematically incorporate explicit native 3D geometric representations and user trajectory into a wireless predictive framework, our WWM provides a crucial reference for future exploration in AI-native 6G and the development of more sophisticated wireless autonomous agents. The main contributions are as follows.
\begin{itemize}
\item We build a hybrid multi-modal wireless dataset that combines high-fidelity ray tracing with real 6G-prototype measurements. The dataset spans five urban environments, multiple base-station placements, diverse user speeds, and synchronized CSI, point-cloud, and trajectory observations.
\item We propose WWM, a multi-modal JEPA-based world model with shared cross-modal attention and modality-specific feed-forward experts. The model is designed to learn the physics-informed relationship between scene geometry, user motion, and electromagnetic propagation.
\item We show that one pre-trained WWM backbone can support five downstream radio access network tasks through lightweight task heads, covering simulation and real-measurement settings.
\item We demonstrate that multi-modal pre-training improves in-distribution task accuracy, unseen-city generalization, and real-world adaptation.
\end{itemize}
The remainder of this paper is organized as follows. Section II overviews the large-scale hybrid multi-modal dataset we constructed to train and test the proposed Wireless World Model. Section III describes the proposed WWM architecture. Pre-training methods and configurations are described in Section IV. Evaluation results are shown and discussed in Section V. We conclude the paper in Section VI.

\section{Large-scale Hybrid multi-modal dataset}
\label{sec:dataset}

\begin{figure*}[t]
\centering
\includegraphics[width=0.8\textwidth]{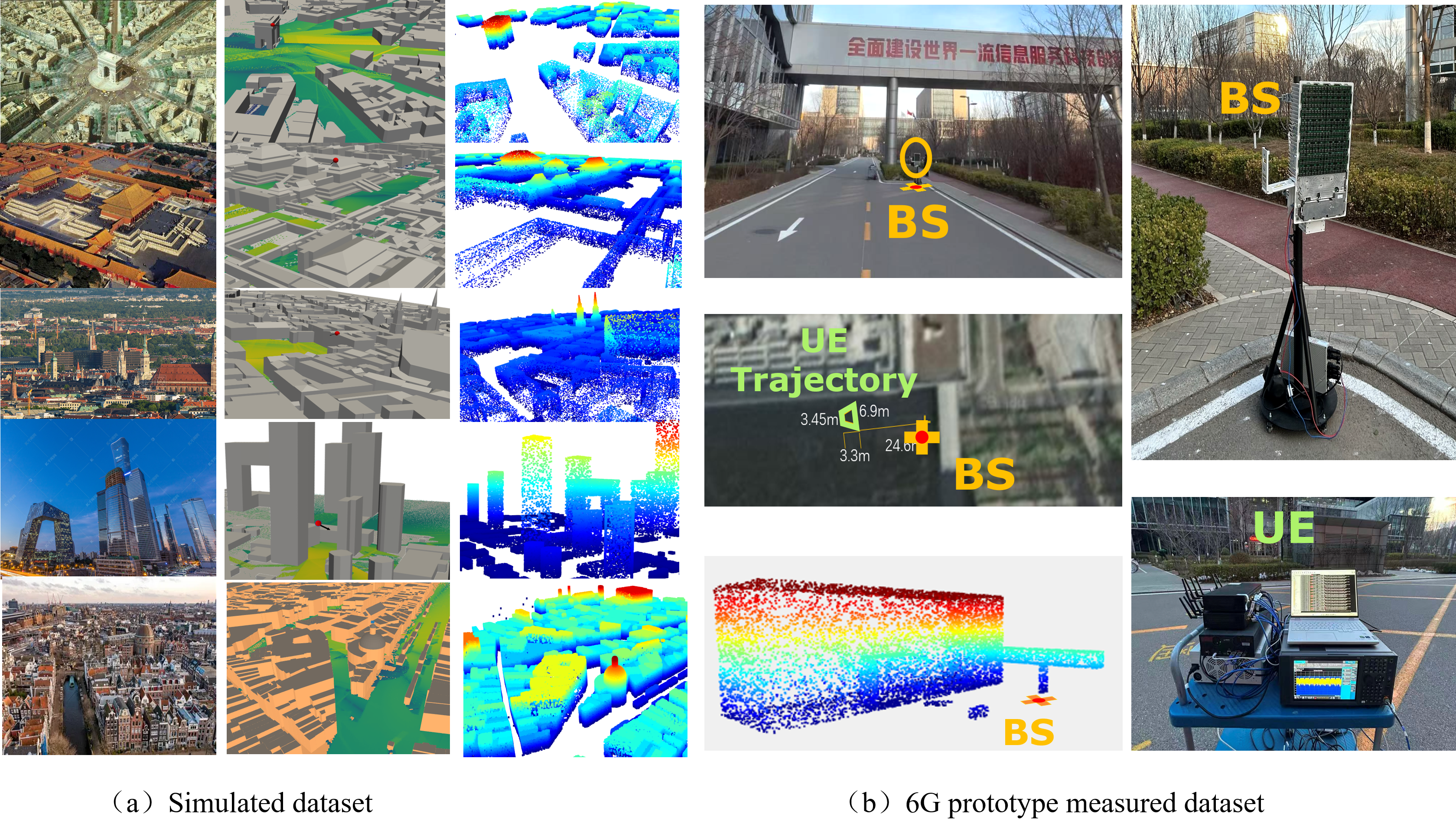}
\caption{Hybrid multi-modal dataset used for pre-training and evaluation. The simulated dataset spans multiple real urban layouts, while the real-world dataset provides a real deployment environment with synchronized CSI, point clouds, and trajectory information.}
\label{fig:dataset}
\end{figure*}


To establish a comprehensive and realistic data foundation for the WWM, we constructed a large-scale multi-modal dataset integrating urban geometric information, wireless channel state information (CSI), and user trajectories, alongside a complementary real-world dataset to validate model generalization under empirical conditions, as shown in Fig. \ref{fig:dataset}. 
\subsection{Simulated dataset}
Five representative urban environments were selected to capture diverse city structure: the Forbidden City (Beijing), the central business district (Beijing), Place de l'Étoile (Paris), an urban district of Munich and Amsterdam. Raw geographic data were obtained from OpenStreetMap, which provides building footprints, road layouts and associated metadata at city scale~\cite{osm}. These map data were converted into three-dimensional urban scenes through a preprocessing pipeline in which building footprints were extruded using available height information or standardized urban assumptions when height data were unavailable. The resulting geometries were imported into Blender to refine scene structure and assign electromagnetic surface materials. Since OpenStreetMap-derived meshes may contain non-manifold edges, self-intersections and open surfaces, all scenario elements were further processed using Mitsuba~\cite{mitsuba} to generate closed manifold triangular meshes. This procedure resolves non-manifold geometry, unifies surface normals and removes degenerate triangles, ensuring geometric consistency for ray-based electromagnetic simulation. The sanitized scenes were exported in a ray-tracing-compatible format that serves as a unified geometric backend for both wireless channel simulation and environment point-cloud generation.

Wireless channels were generated using the physics-based ray tracing framework provided by Sionna~\cite{sionna}. For each scenario, multiple base-station placements were simulated, with CSI computed along continuous UE trajectories spanning several city blocks. The UE trajectories follow straight-line paths at constant speeds typical of normal pedestrians and vehicles. At regularly sampled time instants along each trajectory, frequency-selective MIMO channels were computed based on line-of-sight and specular reflection propagation paths determined by the urban geometry and material properties.

Each channel realization is represented as a complex-valued tensor in the form
$
\mathbb{C}^{N_{\mathrm{UE}} \times N_r \times N_t \times T \times F},
$
where $N_{\mathrm{UE}}$ denotes the number of UE samples, $N_r$ and $N_t$ are the numbers of receive and transmit antenna ports, $T$ is the number of timesteps per sample and $F$ is the number of subcarriers. The antenna dimensions $N_t$ and $N_r$ are determined by the antenna array size and polarization configuration at the base station and UE. Simulation parameters and their correspondence to these variables are summarized in Table \ref{tab:dataset_configuration}.

To facilitate storage and model training, the complex CSI tensors were transformed into a real-valued representation by separating real and imaginary components and grouping subcarriers into non-overlapping subbands. The resulting representation has the form
$
\mathbb{R}^{N_{\mathrm{UE}} \times 2 \times T \times N_t \times N_r'},$ where the factor of two corresponds to real and imaginary components and $N_r' = N_r \times N_{\mathrm{sb}}$, with $N_{\mathrm{sb}}$ denoting the number of subbands obtained by aggregating consecutive 12 subcarriers. This representation preserves frequency selectivity at the subband level while reducing the dimensionality of the channel tensor.

In parallel with channel simulation, explicit geometric representations of the physical environment were constructed as 3D point clouds. Using the same scene descriptions employed for ray tracing, points were sampled from the surfaces of all triangular meshes, producing global-view point clouds represented as $\mathbb{R}^{N_{\mathrm{pts}} \times 3}$ that encode the static geometry of each urban environment, where $N_{\mathrm{pts}}$ denotes the number of point clouds in the environment.

All modalities—including CSI, 3D point clouds and User trajectories—are expressed in a unified global coordinate system and temporally synchronized. The final pre-training dataset contains 24 simulated subsets spanning four urban environments (Forbidden City, Beijing CBD, Munich and Place de l’Étoile), multiple base-station deployments and UE speeds of 5, 30 and 60 km/h. To evaluate generalization, the Amsterdam datasets (5, 30 and 60 km/h) were excluded from pre-training and used exclusively for scenario-level testing. In addition, Beijing CBD datasets collected at previously unseen speeds of 40 and 70 km/h were reserved for speed-level generalization evaluation.

\subsection{Real-world dataset}
To evaluate model performance under empirical wireless conditions, we collected a real-world wireless dataset containing uplink CSI measurements derived from SRS captured using a 6G prototype system developed by the China Mobile Research Institute. The prototype platform integrates advanced transmission technologies and supports high-fidelity wireless experimentation and data acquisition. We collected outdoor wireless dataset at the China Mobile International Information Port in Beijing using a carrier frequency of 6.6\,GHz and a bandwidth of 400\,MHz. The resulting dataset provides realistic channel observations that include hardware impairments, environmental noise and non-ideal propagation effects. The 3D point clouds and user trajectories were constructed from on-site geospatial measurements. Detailed system parameters are summarized in Table \ref{tab:dataset_configuration}.




\begin{table}[!h]
\centering
\caption{Hybrid Large-scale multi-modal dataset configuration}
\scriptsize
\setlength{\tabcolsep}{2pt}
\renewcommand{\arraystretch}{1.05}
\begin{tabularx}{\columnwidth}{@{}l l X X@{}}
\hline
Symbol & Description & Simulation & Real measured \\
\hline
$f_c$ & Center freq. & 2.6~GHz & 6.6~GHz \\
$\Delta f$ & Subcarrier spacing & 15~kHz & 120~kHz \\
$F$ & Subcarriers & 96 & 3333 \\
$N_{\mathrm{sb}}$ & Subbands & 8 & -- \\
$F_{\mathrm{sb}}$ & Subcarriers per subband & 12 & -- \\
$N_{\mathrm{RB}}$ & PRBs & -- & 264 \\
$N_{\mathrm{RB}sp}$ & PRBs/sample & -- & 8 \\
$\Delta t$ & Temporal interval & 5~ms & 10~ms \\
$T$ & Temporal samples & 16 & 16 \\
$N_t$ & BS antenna ports &
$32=4{\times}4{\times}2$ &
$32=16{\times}1{\times}2$ \\
$N_r$ & UE antenna ports &
$4=2{\times}1{\times}2$ &
$4=4{\times}1{\times}1$ \\
$N_r'$ & Effective RX-freq. dim. &
$32=N_r N_{\mathrm{sb}}$ &
$32=N_r N_{\mathrm{RB}sp}$ \\
$d_{\mathrm{ant}}$ & Antenna spacing & $0.5\lambda$ & $0.5\lambda$ \\
$N_{\mathrm{scen}}$ & Urban scenarios & 5 & -- \\
\hline
\label{tab:dataset_configuration}
\end{tabularx}
\end{table}




\section{Wireless World Model Architecture}
\label{sec:model}

\begin{figure*}[!t]
    \centering
    \includegraphics[width=1\linewidth]{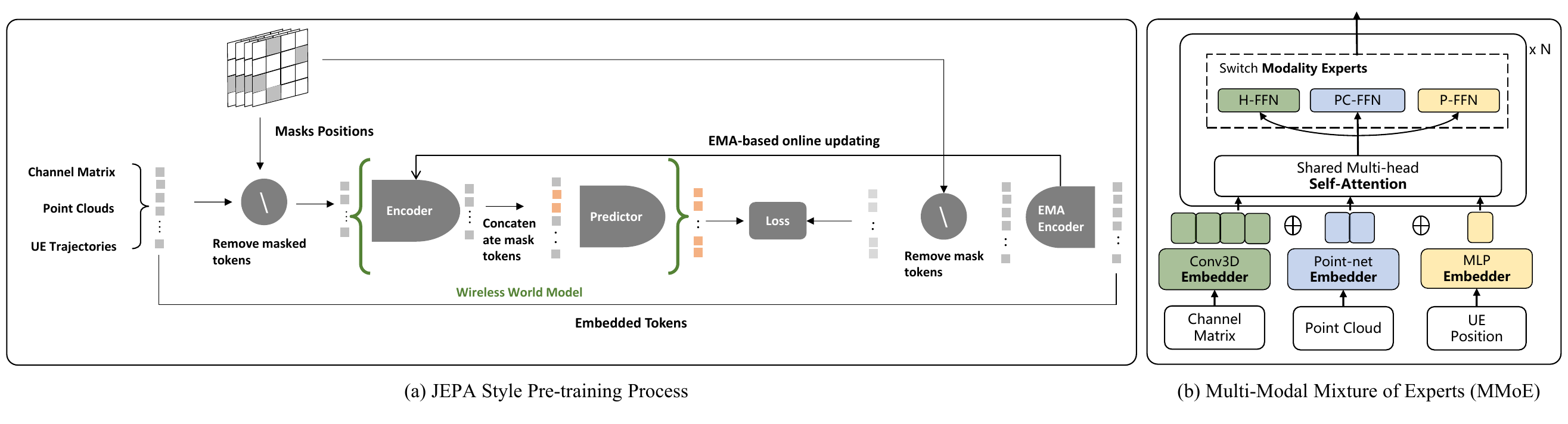}
    \caption{(a) WWM employs a Joint Embedding Predictive Architecture (JEPA) to infer masked multi-modal features in latent space. An online encoder processes visible tokens while a predictor estimates masked embeddings, supervised by an Exponential Moving Average (EMA) based momentum encoder to ensure representation stability. (b) Multi-modal Mixture of Experts (MMoE). Heterogeneous inputs—CSI, 3D point clouds, and trajectories—are tokenized via domain-specific embedders (Conv3D, Point-net, and MLP). Within each Transformer block, shared self-attention performs global cross-modal reasoning, followed by modality-specific experts (H-FFN, PC-FFN, P-FFN) to preserve physical inductive biases.}
    \label{fig:pre-training}
\end{figure*}

The WWM is implemented as a multi-modal JEPA. As shown in Fig. \ref{fig:pre-training} a, the model comprises three main components: an online encoder $f_{\theta}$, a predictor $g_{\phi}$ and a target encoder $f_{\bar{\theta}}$ whose parameters are maintained as an exponential moving average (EMA) of the online encoder. The online encoder operates on partially observed inputs and produces context embeddings, the predictor uses these embeddings together with learnable mask tokens to infer the representations of masked regions, and the target encoder provides slowly varying target embeddings for the same inputs without masking. All supervision is applied in the latent space, and the model is never asked to reconstruct raw samples.

\subsection{Multi-modal Joint Embedding and Prediction Architecture}

For each training sample, the raw input consists of three synchronized modalities:
a CSI tensor $x^{\mathrm{CSI}}$, a 3D point cloud tensor $x^{\mathrm{PC}}$ and a user trajectory vector $x^{\mathrm{POS}}$ in a BS-centered coordinate frame. After modality-specific embedding (further explained in Section III.B), the resulting token sets $\mathbf{X}^{\mathrm{CSI}},\, \mathbf{X}^{\mathrm{PC}},\, \mathbf{X}^{\mathrm{POS}}$are concatenated into a unified sequence
$\mathbf{X}_0 = \mathrm{Concat}\!\big( \mathbf{X}^{\mathrm{CSI}} \,,\, \mathbf{X}^{\mathrm{PC}} \,,\, \mathbf{X}^{\mathrm{POS}} \big) \in \mathbb{R}^{N \times D}$. 
A masking module then selects two disjoint index sets over this sequence,

\begin{equation}
\mathcal{I}_{\text{full}} = \mathcal{I}_{\text{enc}} \sqcup \mathcal{I}_{\text{pred}},
\end{equation}
which specify visible (context) tokens $\mathcal{I}_{\mathrm{enc}}$ and masked (prediction) tokens $\mathcal{I}_{\mathrm{pred}}$ indices, respectively. The online encoder $f_{\theta}$ receives only the visible tokens and produces context embeddings
\begin{equation}
z_{\mathrm{ctx}} = f_{\theta}(X; \mathcal{I}_{\mathrm{enc}}),
\end{equation}
In parallel, the target encoder $f_{\bar{\theta}}$ processes the complete, unmasked input and outputs target embeddings
\begin{equation}
h = f_{\bar{\theta}}(X; \mathcal{I}_{\mathrm{full}}),
\end{equation}
from which the targets at the masked positions $h_{\mathcal{I}_{\mathrm{pred}}}$ are extracted. The predictor $g_{\phi}$ then takes the context embeddings together with a set of learnable mask tokens $\{\mathbf{m}_j \in \mathbb{R}^D\}_{j \in \mathcal{I}_{\mathrm{pred}}}$, each associated with a masked position in $\mathcal{I}_{\mathrm{pred}}$, and produces predicted embeddings
\begin{equation}
\hat{h}_{\mathcal{I}_{\mathrm{pred}}}
= g_{\phi}\big(z_{\mathrm{ctx}},\, \{\mathbf{m}_j\}_{j \in \mathcal{I}_{\mathrm{pred}}};\, \mathcal{I}_{\mathrm{pred}}\big),
\end{equation}
A latent-space loss (here an $\ell_1$ distance) is computed between $\hat{h}_{\mathcal{I}_{\mathrm{pred}}}$ and $h_{\mathcal{I}_{\mathrm{pred}}}$, and gradients are applied to $\theta$ and $\phi$ only; the target parameters $\bar{\theta}$ are updated via EMA. This scheme encourages the model to learn stable, semantically meaningful embeddings that capture the physical structure linking channel responses, geometry and motion.

\subsection{Multi-modal Input Fusion}\label{embeddings}

To enable joint processing of heterogeneous modalities, all inputs are mapped to a shared embedding space with dimension $D$, as shown in Fig. \ref{fig:pre-training} b. Each modality is first converted into a sequence of tokens using a modality-specific encoder, and these tokens are then projected into the common latent dimension.

\paragraph{Channel (CSI) embedding}

The raw CSI tensor $x^{\mathrm{CSI}} \in \mathbb{R}^{C_{\mathrm{in}} \times T \times H \times W}$
(where $C_{\mathrm{in}}=2$ corresponds to the real and imaginary components) is treated as a 3D spatiotemporal volume over time, frequency, and antenna (spatial) indices. A 3D convolution with kernel size and stride both equal to $(T_p, H_p, W_p)$ serves as the patchification operator: it partitions the volume into non-overlapping tubes of size $(T_p, H_p, W_p)$ and simultaneously projects each tube into a
$D$-dimensional embedding vector, yielding $N_{\mathrm{CSI}} = \frac{T}{T_p} \times \frac{H}{H_p} \times \frac{W}{W_p}$ CSI tokens $\{\mathbf{X}^{\mathrm{CSI}}_i \in \mathbb{R}^D\}_{i=1}^{N_{\mathrm{CSI}}}$.
To preserve the spatiotemporal ordering, a fixed 3D sinusoidal positional encoding is added to each token.

\paragraph{Point-cloud embedding}

The raw 3D point cloud $x^{\mathrm{PC}} \in \mathbb{R}^{N_{\mathrm{pts}} \times 3}$ is encoded using a discrete-variational tokenizer inspired by point-cloud auto-encoding methods. Farthest-point sampling selects a fixed number of centers, and local neighborhoods around these centers are constructed by nearest-neighbor grouping. A lightweight PointNet-style encoder from Point-BERT \cite{yu2022point} extracts a feature vector for each neighborhood, which is then passed through a learned code book via Gumbel–Softmax quantization and refined by a shallow geometric network. The result is a set of $N_{\mathrm{PC}}$ patch-level point cloud tokens in the shared latent space, $x^{\mathrm{PC}} \rightarrow \{ \mathbf{X}^{\mathrm{PC}}_j \in \mathbb{R}^D \}_{j=1}^{N_{\mathrm{PC}}}$. This procedure preserves local geometry while compressing the raw point cloud into a compact, fixed-size sequence.

\paragraph{Trajectory embedding}

The raw user trajectory vector $x^{\mathrm{POS}} = \{\mathbf{p}_t \in \mathbb{R}^3\}_{t=1}^{T_{\mathrm{pos}}}$ is represented as a sequence of positions over time. A small projection network, implemented as a multi-layer perceptron with non-linear activations, maps each position to a $D$-dimensional embedding, and a temporal positional encoding is added: $x^{\mathrm{POS}} \rightarrow \{ \mathbf{X}^{\mathrm{POS}}_k \in \mathbb{R}^D \}_{k=1}^{N_{\mathrm{POS}}}$.

\paragraph{Unified token sequence}

After modality-specific embedding, the three token sequences are concatenated along the sequence dimension in a fixed order,
\begin{equation} \mathbf{X}_0 = \mathrm{Concat}\!\big( \mathbf{X}^{\mathrm{CSI}} \,,\, \mathbf{X}^{\mathrm{PC}} \,,\, \mathbf{X}^{\mathrm{POS}} \big) \in \mathbb{R}^{N \times D}, \end{equation}
where
$N = N_{\mathrm{CSI}} + N_{\mathrm{PC}} + N_{\mathrm{POS}}$ and
$\mathbf{X}^{\mathrm{CSI}}$, $\mathbf{X}^{\mathrm{PC}}$, $\mathbf{X}^{\mathrm{POS}}$ denote all the tokens for each modality. The model keeps track of the segment lengths $(N_{\mathrm{CSI}}, N_{\mathrm{PC}}, N_{\mathrm{POS}})$ for use in the modality-aware expert layers. In our implementation, the numbers of tokens for CSI, point cloud, and trajectory are $N_{\mathrm{CSI}}=512$, $N_{\mathrm{PC}}=256$, and $N_{\mathrm{POS}}=16$, respectively.

\subsection{Multi-modal Mixture of Expert Block}

The unified token sequence $\mathbf{X}_0$ is processed by a stack of $L_e$ Transformer blocks. As illustrated in Fig. \ref{fig:pre-training} b, within each block, layer-normalized tokens first pass through a shared multi-head self-attention (MHSA) layer that operates on all modalities jointly, enabling cross-modal information exchange. The output is then layer-normalized, split by modality and routed to three parallel feed-forward sub-networks, each specialized to CSI, point cloud, or trajectory tokens respectively. The expert outputs are concatenated back and added as a residual. This two-stage process repeats for $L_e$ layers, yielding the final representation $\mathbf{X}_{L_e}$.

\paragraph{Shared self-attention}

Given the input $\mathbf{X}_{\ell} \in \mathbb{R}^{N \times D}$ at layer $\ell$, a standard multi-head self-attention (MHSA) layer with pre-normalization and residual connection is applied:
\begin{equation}
\mathbf{X}'_{\ell}
= \mathbf{X}_{\ell}
  + \mathrm{MHSA}\big(\mathrm{LN}(\mathbf{X}_{\ell})\big).
\end{equation}
Because $\mathbf{X}_{\ell}$ contains tokens from all three modalities, the attention mechanism learns to exchange information across CSI, geometry and motion, for example by allowing channel tokens to attend to nearby building structures or trajectory tokens.

\paragraph{Modality-specific feed-forward experts}

Rather than a single shared feed-forward layer, 3 feed-forward sub-network, one for each modality, in each Transformer block is implemented as three parallel experts. After a second normalization, the sequence $\mathbf{X}'_{\ell}$ is partitioned into CSI, 3D point cloud and user trajectory segments according to the stored lengths:
\begin{equation}
\big[
\mathbf{X}^{\mathrm{CSI}}_{\ell},
\mathbf{X}^{\mathrm{PC}}_{\ell},
\mathbf{X}^{\mathrm{POS}}_{\ell}
\big]
=
\mathrm{Split}\!\left(
    \mathrm{LN}(\mathbf{X}'_{\ell});
    N_{\mathrm{CSI}}, N_{\mathrm{PC}}, N_{\mathrm{POS}}
\right).
\end{equation}
Each segment is then passed through its own two-layer feed-forward expert,
\begin{equation}
\begin{aligned}
\tilde{\mathbf{X}}^{\mathrm{CSI}}_{\ell} &= f_{\mathrm{CSI}}\!\big(\mathbf{X}^{\mathrm{CSI}}_{\ell}\big), \\
\tilde{\mathbf{X}}^{\mathrm{PC}}_{\ell}  &= f_{\mathrm{PC}}\!\big(\mathbf{X}^{\mathrm{PC}}_{\ell}\big), \\
\tilde{\mathbf{X}}^{\mathrm{POS}}_{\ell} &= f_{\mathrm{POS}}\!\big(\mathbf{X}^{\mathrm{POS}}_{\ell}\big).
\end{aligned}
\end{equation}
where each feed-forward expert $f$ is a position-wise non-linear mapping with separate parameters. The updated segments are concatenated back to their original order and combined with a residual connection:
\begin{equation}
\mathbf{X}_{\ell+1}
= \mathbf{X}'_{\ell}
+ \mathrm{Concat}\!\big(
    \tilde{\mathbf{X}}^{\mathrm{CSI}}_{\ell},
    \tilde{\mathbf{X}}^{\mathrm{PC}}_{\ell},
    \tilde{\mathbf{X}}^{\mathrm{POS}}_{\ell}
  \big).
\end{equation}
This ``shared-attention plus modality-expert'' design can be viewed as a multi-modal mixture-of-experts architecture with deterministic routing based on modality identity. It allows global context modeling to be shared across modalities, while maintaining specialized pathways tuned to the statistics and physical constraints of CSI, 3D point cloud, and user trajectory.

\subsection{Encoder and predictor configurations}

Both the online encoder $f_{\theta}$ and the predictor $g_{\phi}$ are built from the shared-attention plus modality-expert Transformer blocks described in the previous section, and share the same ViT-Small hyper-parameters ($D{=}384$, $L_e{=}L_p{=}12$). Despite this architectural symmetry, the two networks serve distinct roles: the encoder operates directly on the embedded input tokens at the visible positions $\mathcal{I}_{\mathrm{enc}}$ and must learn rich, general-purpose representations of the observed context; the predictor, by contrast, receives the encoder's output together with learnable mask tokens at positions $\mathcal{I}_{\mathrm{pred}}$ and is tasked with inferring the latent content of the masked regions. The encoder and predictor adopt a similar architectural design, each comprising 12 MMoE Transformer layers. The embedding dimension is set to 384; the encoder uses 6 attention heads with a head dimension of 64, while the predictor employs 12 attention heads with a head dimension of 32. Together, the encoder and predictor have a total of 78 million parameters.

\section{WWM Pre-training}
The WWM was pre-trained in a self-supervised manner using JEPA described above. In this setting, the model is presented with partially observed multi-modal inputs and is trained to predict the latent representations of the masked regions in a shared embedding space, rather than reconstructing raw measurements. This approach encourages the encoder--predictor pair to capture the underlying physical structure of the wireless environment.

\subsection{Data preprocessing}
\label{sec:data_preprocessing}

Raw CSI tensors acquired from either simulation or the 6G prototype system are preprocessed into a numerically stable representation to facilitate large-scale self-supervised training.

Samples containing zero-valued elements are first discarded. The remaining CSI values span a wide dynamic range, which can hinder training stability. To compress this range while preserving sign information, we apply a signed log transform:
\begin{equation}
\tilde{\mathbf{H}} = \mathrm{sign}(\mathbf{H}) \cdot \log\!\big(1 + |\mathbf{H}|/\epsilon\big),
\end{equation}
where $\mathbf{H}$ denotes the CSI tensor and $\epsilon$ is a small scaling constant that amplifies near-zero magnitudes before the logarithm, ensuring fine-grained distinctions among weak signal components are retained. We set $\epsilon = 10^{-7}$ for simulated data and $\epsilon = 10^{-2}$ for 6G prototyple measured data. Finally, mean--variance standardization is applied to rescale the dataset to zero mean and unit standard deviation. This pipeline converts heterogeneous raw inputs into clean, standardized tensors with controlled dynamic range, which is critical for stable JEPA pre-training across diverse cities and user speeds.

\subsection{Masking strategies for CSI and trajectories}

To expose the model to complementary forms of partial observability, three masking configurations were interleaved during pre-training. All configurations operate on the unified token sequence obtained by concatenating CSI, point-cloud and trajectory tokens, but place the emphasis on different aspects of the wireless scenario.

\textbf{Fine-grained CSI masking}: In the first configuration, the CSI volume is partitioned into a 3D grid of spatiotemporal patches along time, frequency and antenna (or spatial) dimensions. For each sample, several relatively small 3D blocks are sampled at random within this grid, and all CSI patches inside these blocks are designated as masked. Concretely, we sample $8$ blocks per clip, each with a temporal extent covering $50\%$ of the tubeletized time axis and a spatial extent corresponding to approximately $15\%$ of the CSI patch grid in both spatial dimensions. The remaining CSI patches, together with all point-cloud and trajectory tokens, form the visible context. This configuration yields a moderate masking ratio over CSI and encourages the model to reconstruct fine-scale multipath structure when sufficient local context is available.

\textbf{Coarse CSI masking}: In the second configuration, fewer but substantially larger 3D blocks are masked in the CSI grid, producing sizeable ``holes'' in time–frequency–space that must be inferred from the remaining CSI context. Here we mask $2$ blocks per clip, each covering $50\%$ of the tubeletized time axis and roughly $70\%$ of the patch grid in each spatial dimension. Point-cloud and trajectory tokens remain fully visible. Compared with the fine-grained configuration, this setting places more emphasis on long-range dependencies and global consistency.

\textbf{Trajectory masking}: The third configuration targets user trajectory inference. In this setting, all CSI patches and point-cloud tokens remain fully visible. Instead, the entire trajectory token sequence is masked. The model thus receives a complete description of the channel evolution and environment, but no explicit user coordinates, and is required to reconstruct the latent embeddings associated with the trajectory. This configuration encourages the model to internalize the inverse relationship from channel and geometry back to user motion patterns. 

Across a mini-batch, the three configurations are sampled with equal probability, and the total loss is obtained by averaging the latent-space prediction losses from each configuration. This multi-task JEPA objective ensures that the same pre-trained model is simultaneously optimized for both channel completion and trajectory inference under different visibility patterns.

\subsection{Pre-training configurations}

The WWM model, including encoder and predictor, was pre-trained on a simulated large-scale multi-modal dataset, as described in Section II.A, with each sample containing CSI tokens of 16 time steps, 3D point cloud tokens and synchronized user trajectory vector. For each sample, the input point cloud is cropped from the global scene using a 3D bounding cuboid defined by the UE and serving-BS locations, with the cuboid enlarged by a factor of two to retain surrounding propagation-relevant geometry. The model was trained with a global batch size of $128$ using the AdamW optimizer and the L1 loss function. Following a cosine learning-rate schedule, the learning rate was linearly warmed up from $1.0\times 10^{-5}$ to a peak value of $2.0\times 10^{-5}$ over the first $2$ epochs, and then decayed by a cosine scheduler to a final learning rate of $1.0\times 10^{-5}$ by the end of training. Weight decay was fixed at $0.04$ throughout. The target encoder was updated with a momentum of $0.9925$ (applied to both the encoder and predictor branches), providing a slowly evolving target network that stabilizes training in the JEPA setting. Pre-training was run for 16 epochs with a mixed-precision bfloat16 data type. Notably, the full pre-training workflow on the complete multi-modal dataset was completed in 87 hours on a single NVIDIA RTX 4090 24GB consumer-grade GPU, demonstrating an extremely low training computational barrier compared to large-scale foundation models in other domains. 

\section{Downstream task adaptation}
\subsection{CSI Temporal Prediction}
In CSI time-domain prediction task, the WWM encoder and predictor are configured to predict future CSI in the latent space based on 14 historical CSI timesteps, the corresponding 14 historical user positions, and the 3D point cloud, while a decoder is trained to recover CSI from the latent representation.

In this setting, the WWM encoder and predictor operate exactly as in the pre-training stage, but their parameters are kept frozen. A multi-time step CSI sample, along with the associated point cloud and user trajectory, is fed into the encoder. Only the first 14 time steps in the sample are treated as visible context, while the channel tokens corresponding to subsequent $T_{\mathrm{pred}}$ = 2 time steps within the sample are designated as masked positions. The predictor receives the context embeddings and a set of learnable mask tokens at these future positions, and produces latent tokens that represent the predicted CSI in the embedding space. These predicted tokens, restricted to the CSI modality, are then passed to a dedicated channel decoder that maps them back to the complex CSI tensor on the original time--frequency--space format.

The channel decoder is implemented as a compact transformer-based network specialized for CSI reconstruction. It consists of a stack of Transformer blocks operating purely on CSI tokens, followed by a projection head that reshapes the output into a tensor of shape $(2, T_{\mathrm{pred}}, H, W)$. Here, the first dimension corresponds to real and imaginary parts; $T_{\mathrm{pred}}$ denotes the number of future channel time steps being predicted, $H$ corresponds to the number of base-station antennas (32 in our dataset), and $W$ denotes the joint frequency--antenna dimension defined by the product of user-side antenna elements and subband groups (here $4 \times 8$). This structure allows the decoder to reconstruct the full complex CSI on time--frequency--space format from the predicted latent tokens. In our implementation, the decoder is a 6-layer transformer trained with AdamW and a learning rate of $2 \times 10^{-5}$.

The decoder is trained with a complex-valued reconstruction loss that explicitly separates magnitude and phase components of the channel, while also incorporating a structural similarity objective. Given predicted and ground-truth channels in the form $\hat{\mathbf{Y}}, \mathbf{Y} \in \mathbb{R}^{2 \times T_{\mathrm{pred}} \times H \times W}$ (real and imaginary parts separated in the first dimension), we form complex tensors $\hat{\mathbf{H}}, \mathbf{H} \in \mathbb{C}^{T_{\mathrm{pred}} \times H \times W}$ and their magnitudes and phases of each element. The CSI loss combines a raw mean-squared error in the real--imaginary plane, a magnitude loss, a phase-consistency term, and an SGCS regularization term:
\begin{equation}
\label{eq:csi_loss}
\begin{aligned}
\mathcal{L}_{\mathrm{CSI}}
=&
\underbrace{\frac{1}{N}\big\|\hat{\mathbf{Y}} - \mathbf{Y}\big\|_2^2}_{\text{raw MSE}}
+ \alpha
\underbrace{\frac{1}{N}\big\||\hat{\mathbf{H}}| - |\mathbf{H}|\big\|_2^2}_{\text{magnitude loss}} \\
&+ \beta
\underbrace{\left(1 - \frac{1}{N}\sum_{n} \cos(\hat{\phi}_n - \phi_n)\right)}_{\text{phase-consistency loss}} \\
&+ \gamma
\underbrace{\frac{1}{T_\mathrm{pred}}\sum_{t=1}^{T_\mathrm{pred}} \mathcal{L}_{\mathrm{SGCS},t}}_{\text{average SGCS loss}} .
\end{aligned}
\end{equation}

Here $N = 2 \times T \times H \times W$ denotes the number of elements in $\hat{\mathbf{Y}}$, and $\hat{\phi}_n$ and $\phi_n$ represent the predicted and true phases at element $n \leq N$. In implementation, $\mathcal{L}_{\mathrm{SGCS}} = 1-\mathrm{SGCS}$ is computed on the final predicted time step by reshaping the reconstructed channel into subcarrier groups and measuring the similarity between the dominant spatial singular vectors of the predicted and ground-truth CSI. Specifically, for each subcarrier $k$, we reshape the channel matrix as $\mathbf{H}_k \in \mathbb{C}^{N_{\mathrm{ue}} \times N_{\mathrm{bs}}}$ and perform singular value decomposition (SVD):
\begin{equation}
\mathbf{H}_k = \mathbf{U}_k \boldsymbol{\Sigma}_k \mathbf{V}_k^{\mathrm{H}} ,
\end{equation}
where the first right singular vector $\mathbf{v}_k$ (corresponding to the largest singular value) captures the dominant spatial direction. Let $\hat{\mathbf{v}}_k$ and $\mathbf{v}_k$ denote the dominant right singular vectors of the predicted and ground-truth CSI, respectively. The SGCS of a single CSI time step is computed as the average normalized inner product across all $K$ subcarriers:
\begin{equation}
\mathrm{SGCS}
=
\frac{1}{K}
\sum_{k=1}^{K}
\frac{
\left| \hat{\mathbf{v}}_k^{\mathrm{H}} \mathbf{v}_k \right|^2
}{
\|\hat{\mathbf{v}}_k\|_2^2 \, \|\mathbf{v}_k\|_2^2
}.
\end{equation}

In our training setup, the CSI-loss weights are progressively adjusted across epochs to balance stable amplitude reconstruction, phase alignment, and SGCS optimization: epochs 1--10 use $\alpha=1.0$ and $\beta=0.2$; epochs 11--15 use $\alpha=1.0$ and $\beta=0.5$; epochs 16--20 use $\alpha=1.0$ and $\beta=1.0$; and epochs 21--25 use $\alpha=1.0$, $\beta=0.2$, and $\gamma=1.0$. During evaluation, we report SGCS to quantify the structural similarity between the reconstructed and ground-truth CSI across the antenna array and frequency bands. 


\subsection{Channel Compression and Feedback}
In massive MIMO systems, minimizing feedback overhead while ensuring high CSI acquisition accuracy is critical for accurate precoding and, consequently, wireless transmission capacity. Building on the representations learned by WWM, a pair of efficient deep learning-based compressor and decompressor networks are trained to substantially reduce the feedback payload while preserving CSI fidelity. 

In this task, channel compression is formulated as a downstream task operating on the latent representations extracted by the pre-trained WWM. Out of a 16-timestep CSI sample (processed as 8 temporal tubelets), only the 4th tubelet (corresponding to time steps 7 and 8) is kept unmasked and fed into WWM encoder (together with the corresponding 3D point cloud and User trajectory), which  produces $N=64$ continuous CSI latent tokens $Z_{\mathrm{csi}} \in \mathbb{R}^{N \times D}$ in the embedding space, each with a dimension of $D=384$. A dedicated lightweight compression head, implemented as \textbf{CRNetTokens}, is attached on top of these tokens. The compression and quantization process can be formulated as:
\begin{equation}
\label{eq:csi_comp}
Z_{\mathrm{comp}} = \mathcal{Q}_{\mu, b}\left( f_{\mathrm{reduce}}(Z_{\mathrm{csi}}) \right)
\end{equation}
where $f_{\mathrm{reduce}}(\cdot)$ applies a dimensionality reduction layer that shrinks the embedding dimension by a configurable reduction factor $r$ (e.g., $r=96$), followed by a token-mixing stage. The function $\mathcal{Q}_{\mu, b}(\cdot)$ represents a $\mu$-law scalar quantizer that maps the continuous vectors to $2^b$ discrete levels (e.g., $b=4$ bits) and is optimized via a straight-through estimator during training.

A decompressor $f_{\mathrm{expand}}(\cdot)$ subsequently expands the discrete feedback payload back to the original token dimension $D$. These reconstructed tokens are then passed to a channel decoder $\mathcal{D}$---a Vision Transformer jointly fine-tuned with the compressor---to recover the complex channel tensor the original time--frequency--space format:
\begin{equation}
\label{eq:csi_decomp}
\hat{\mathbf{Y}} = \mathcal{D}\left( f_{\mathrm{expand}}(Z_{\mathrm{comp}}) \right)
\end{equation}
The compressor and decoder are trained end-to-end using the AdamW optimizer with a learning rate of $1 \times 10^{-4}$ and a weight decay of $0.01$. The training objective is a complex-valued CSI reconstruction loss that combines a raw mean-squared error, a magnitude loss (weight $\alpha=1.0$), and a phase-consistency term (weight $\beta=0.5$).

\subsection{Beam Prediction}
Sub-6~GHz signals propagate through the same physical environment as upper-6G (U6G) signals and therefore their dominant propagation paths are shaped by the same geometry. Leveraging this property, the BS can use Sub-6~GHz CSI to predict the most suitable U6G beam direction, thereby avoiding an exhaustive pilot-based beam search and reducing both air-interface overhead and energy consumption. A single-layer transformer is trained with WWM encoded Sub-6~GHz CSI to predict the Precoding matrix indicator (PMI) index of a predefined Type I Single-Panel Codebook of 64 beams.

To be more specific, the beam prediction task adopts the same urban scenarios and user velocity ranges as those used for pre-training and other downstream tasks. For each user trajectory, CSI is collected at two distinct central frequency simultaneously: a Sub-6GHz band (2.6GHz) and an upper 6GHz (U6G) band (6.62505GHz). At every sampled UE position, the Sub-6GHz CSI tensor $\mathbf{X}_{\text{sub-6}}$ is used as input, together with the corresponding 3D point cloud and user trajectory. From the concurrent U6G channel, the optimal PMI index $b^*$ is determined as the beam that maximizes the received power over a predefined Type I Single-Panel Codebook of $K$ beams. Thus the task learns the mapping:
\begin{equation}
f_{\text{beam}}:\mathbf{X}_{\text{sub-6}} \;\longrightarrow\; b^{*} \in \{1,2,\dots,K\},
\end{equation}

where $f_{\text{beam}}$ denotes the prediction model and $\mathbf{X}_{\text{sub-6}} \in \mathbb{R}^{2 \times T \times N_t \times N_r'}$ is the processed low-frequency CSI tensor. The mapping is learned using the pre-trained WWM as a frozen feature extractor, followed by a task-specific one-layer attentive classifier. The attentive classifier takes the encoded feature tokens and outputs a probability distribution over the $K$ beams, trained with the cross-entropy loss:

\begin{equation}
\mathcal{L}_{\text{beam}} = -\frac{1}{N} \sum_{i=1}^{N} \sum_{k=1}^{K} y_{i,k} \log(p_{i,k}),
\end{equation}

where $N$ is the batch size, $y_{i,k}$ is the one-hot encoding of the ground-truth beam index for the $i$-th sample, and $p_{i,k}$ is the predicted probability for the $k$-th beam. Throughout training, only the attentive classifier parameters are updated, while the WWM backbone remains fixed.

To assess the prediction performance, we measure the Top-1 classification accuracy, as well as the achieved beam gain relative to the optimal beam. The beam gain ratio for a set of $M$ test samples is computed as:

\begin{equation}
R_{\text{BG}} = 
\frac{1}{M}
\sum_{j=1}^{M}
\frac{
\left\| H_{U6G} w_{b}^{(j)} \right\|_2^2
}{
\left\| H_{U6G} w_{b^{*}}^{(j)} \right\|_2^2
}.
\end{equation}

where $H_{U6G}$ is the last timestep CSI of $j$-th sample in U6G band, $w_{b}^{(j)}$ is the PMI corresponding to predicted beam index for the $j$-th sample, and ${w_{b^{*}}^{(j)}}$ is the PMI corresponding to the theoretical optimal beam. This metric directly reflects how closely the model's beam selections approach the ideal link performance.


\subsection{User Localization}
Leveraging its multi-modal understanding of the environment, WWM enables high-precision user localization by treating wireless CSI as a distinctive ``fingerprint''.
In this downstream task, the model is required to infer the 2D geographical coordinates $(x, y)$ of a UE based on its CSI as well as the corresponding 3D cloud point. Similar to the channel prediction task, the pre-trained WWM backbone remains frozen to leverage its internalized physical representations. The explicit user trajectory tokens are intentionally omitted from WWM input, compelling the model to derive positional information solely from the interaction between CSI patterns and the geometric structure of the environment, where the point cloud provides a coarse-grained representation of the UE–BS propagation environment obtained by cropping the global scene with a 3D bounding cuboid defined by the UE and BS locations and enlarged by a factor of two. The resulting latent tokens from the WWM encoder, which encapsulate joint EM-geometric features, are then fed into a dedicated attentive regression head. The attentive regression head is designed to regress the sequence of latent embeddings into a 2D coordinate $\hat{\mathbf{p}} = (\hat{x}, \hat{y})$. The localization head is trained to minimize the Euclidean distance between the predicted coordinates and the ground-truth position $\mathbf{p} = (x, y)$ corresponding to the final position of the UE trajectory. The objective function is defined by the mean squared error (MSE) loss. We quantify performance using the CDF of the absolute localization error.


\subsection{CSI Frequency-Domain Prediction Based on Real-world SRS Measurement}
In time-division duplex (TDD) systems, uplink and downlink channels are reciprocal within the coherence time, allowing BS precoding and multi-user coordination to rely on uplink channel observations. Sounding reference signals (SRS) provide uplink CSI at the BS; inferring unmeasured frequency resources from partial SRS observations reduces measurement overhead while maintaining channel awareness for precoding, scheduling and link adaptation. We adapt WWM to the CSI frequency-domain prediction task to evaluate the model's capability to reduce uplink measurement overhead under realistic deployment conditions. 

The data are collected from a 6G prototype system in an outdoor slow-mobility scenario. In practical 6G systems, sounding reference signals (SRS) are transmitted over multiple physical resource blocks (PRBs) in frequency domain for uplink channel estimation at the base station. Dense PRB-level measurements, however, incur substantial signaling overhead. This task aims to infer CSI on unmeasured SRS PRBs from partially observed PRBs within the same time window.

The prototype system operates at 6.6 GHz with 400 MHz bandwidth and 120 kHz subcarrier spacing. Each PRB contains 12 subcarriers (1.44 MHz bandwidth). The raw measurements comprise 264 PRBs, which are partitioned into 33 samples of 8 consecutive PRBs. For each sample, the first four PRBs serve as visible context, and the remaining four PRBs are designated as prediction targets. Each sample corresponds to a continuous UE trajectory containing 16 consecutive CSI time steps, which are jointly modeled to capture temporal channel evolution.

The task is formulated as learning the mapping:
\begin{equation}
f_{\text{CSI}}: X_{\text{vis}} \rightarrow X_{\text{mask}},
\end{equation}
where $X_{\text{vis}}$ and $X_{\text{mask}}$ denote the CSI tensors of the observed and masked PRBs, respectively.

CSI frequency-domain prediction is implemented as a downstream task on top of the pre-trained WWM. The encoder and predictor retain the same architecture as in pre-training. In practice, training on the measured dataset is conducted in two stages. First, the encoder and predictor are jointly fine-tuned to adapt the pre-trained representations to real-world data. Then, the encoder is frozen, and the predictor is further optimized together with a lightweight Transformer-based channel decoder.
A full 16-timestep trajectory sample, together with its 3D point cloud and user trajectory tokens, is fed into the encoder. Within each time step, only the first four PRBs are provided as input, while the remaining four PRBs are withheld as prediction targets.
The predictor produces latent representations corresponding to the full CSI matrix, rather than only masked positions. These representations are passed to the channel decoder, which operates solely on CSI tokens and outputs a complete CSI estimate. A final projection head reshapes the output into a tensor of size $(2, T, H, W)$, where the first dimension corresponds to real and imaginary parts; $T=16$ is the number of temporal samples; $H=32$ is the number of base-station antennas; and $W$ denotes the joint frequency-antenna dimension (4 UE antennas $\times$ 8 PRBs).
The model is trained using mean squared error between the predicted and ground-truth CSI over the full frequency band, and performance is evaluated with NMSE and SGCS. By reconstructing unobserved PRBs from partial observations, this task reflects the model's ability to exploit frequency-domain channel correlations in real-world scenarios and provides a practical mechanism for reducing SRS overhead in operational 6G systems. 


\section{Evaluation}
\label{sec:adaptation}
\subsection{Electromagnetic
propagation understanding via pre-training}
\begin{figure*}[!t]
\centering
\includegraphics[width=0.75\textwidth]{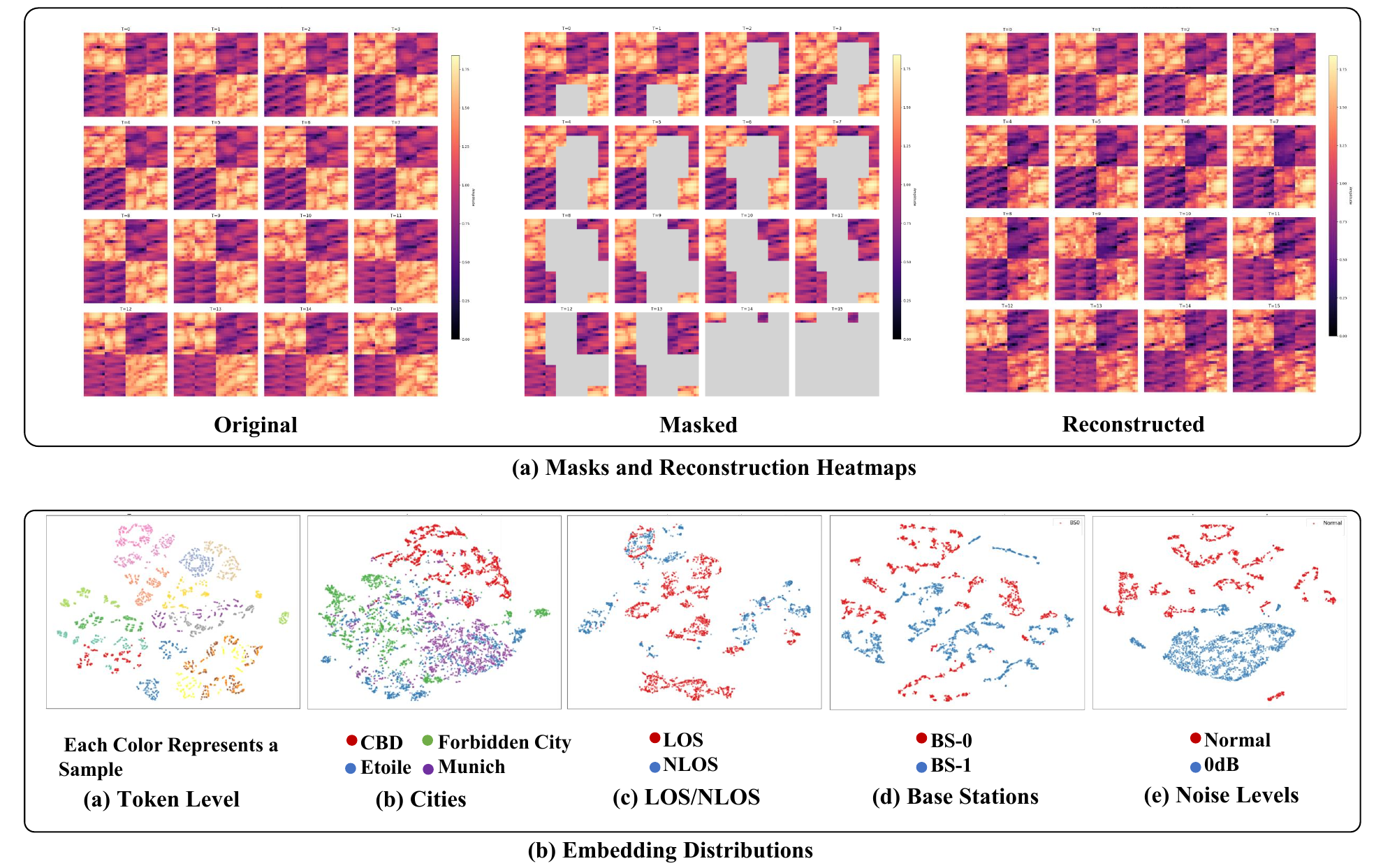}
\caption{(a) Pre-training results, WWM reconstructs masked CSI patterns from multi-modal context. (b)  The t-SNE of WWM encoder’s final-layer embeddings across five labeling aspects}
\label{fig:pretrain_results}
\end{figure*}

The effectiveness of this pre-training is evident in the model's ability to reconstruct missing information from context. Fig. \ref{fig:pretrain_results}(a) visualizes the reconstruction of a masked CSI sample of 16 timesteps. Even when significant time-frequency blocks are masked, WWM accurately restores the channel structure by reasoning from the unmasked CSI blocks, 3D geometry and trajectory cues. We utilized t-distributed stochastic neighbor embedding (t-SNE)~\cite{van2008visualizing} to visualize the WWM encoder's final-layer CSI embeddings with different data label. The outcome, as illustrated in Fig. \ref{fig:pretrain_results}(b), revealed that the embeddings produced by WWM organizes samples into meaningful clusters based on unique characteristics, demonstrating that it has successfully internalized a structured representation of the wireless physical environment without explicit supervision.

\subsection{WWM augments RAN downstream tasks}

\begin{table*}[!h]
    \centering
    \caption{CSI temporal prediction performance of WWM compared to other baselines across in-distribution and generalization scenarios.}
    \begin{tabular}{lccc|ccc|ccc}
        \hline
        \multirow{2}{*}{Scenario} &
        \multicolumn{3}{c|}{WWM} &
        \multicolumn{3}{c|}{WiFo} &
        \multicolumn{3}{c}{LSTM} \\
        & T=15 & T=16 & Avg. & T=15 & T=16 & Avg. & T=15 & T=16 & Avg. \\
        \hline
        CBD
        & 0.707 & 0.741 & \textbf{0.724}
        & 0.613 & 0.600 & 0.606
        & 0.550 & 0.514 & 0.532 \\

        Etoile
        & 0.931 & 0.919 & \textbf{0.925}
        & 0.773 & 0.760 & 0.766
        & 0.808 & 0.791 & 0.799 \\

        Forbidden City
        & 0.943 & 0.941 & \textbf{0.942}
        & 0.792 & 0.786 & 0.789
        & 0.762 & 0.747 & 0.755 \\

        Munich
        & 0.896 & 0.881 & \textbf{0.888}
        & 0.631 & 0.619 & 0.625
        & 0.773 & 0.776 & 0.774 \\

        Velocity Generalization
        & 0.659 & 0.688 & \textbf{0.673}
        & 0.532 & 0.523 & 0.527
        & 0.540 & 0.503 & 0.522 \\

        City Generalization
        & 0.854 & 0.829 & \textbf{0.841}
        & 0.748 & 0.737 & 0.742
        & 0.598 & 0.574 & 0.586 \\
        \hline
    \end{tabular}

    \vspace{4pt}
    \begin{minipage}{0.78\textwidth}
        \raggedright
        \footnotesize
        The SGCS of the predicted CSI is reported. Higher values indicate better prediction performance.
    \end{minipage}
    \label{tab:csi_prediction_comparison}
\end{table*}

We evaluated the performance of the pre-trained WWM across 4 downstream tasks, benchmarking it against SOTA task-specific models and representative open-source WFMs. specifically, WiFo \cite{wifo1} is a channel foundation model that is specially designed for channel prediction tasks, while LWM \cite{lwm} is a static channel foundation model that specialize at beam prediction and LOS/NLOS classification. To ensure a rigorous comparison, WFMs were assessed using either official checkpoints or checkpoints pre-trained on the same multi-modal dataset if training code is provided. We kept backbone frozen for all foundation models, where task-specific knowledge was captured by training lightweight output heads. In contrast, task-specific SOTA baselines were trained full-shot from scratch using the full labeled dataset for each respective task. As detailed in the following sections, WWM consistently achieved SOTA performance across all four tasks. This demonstrates the superior adaptation and transferability of WWM’s latent representations. Furthermore, ablation experiments reveal that reverting WWM to a single-modal configuration—by pre-training without 3D point clouds and user trajectory priors—observes an obvious performance degradation. This underscores the necessity of multi-modal environmental grounding for robust wireless representation.

\begin{figure}[h]
\centering
\includegraphics[width=\columnwidth]{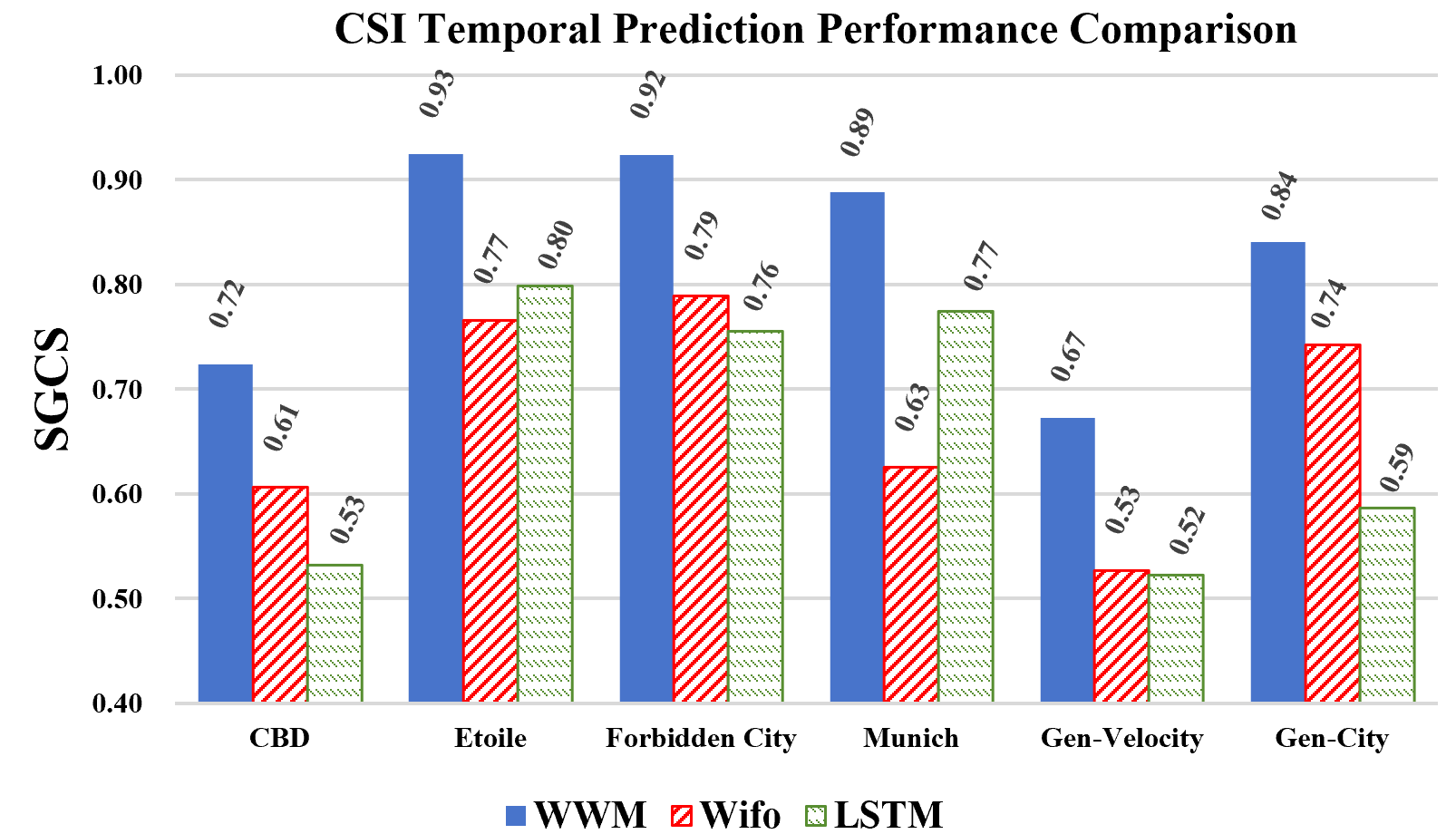}
\caption{CSI Temporal Prediction Performance Comparison}
\label{fig:pred}
\end{figure}

\textbf{CSI temporal prediction}: To assess whether WWM captures channel evolution beyond statistical correlation, we evaluated its performance on CSI temporal prediction against the SOTA baselines WiFo~\cite{wifo1} and LSTM~\cite{graves2012long}, as shown in Fig. \ref{fig:pred}. For in-pattern urban environments (CBD, Etoile, Forbidden City, and Munich), WWM achieves consistently strong Squared Generalized Cosine Similarity (SGCS) scores ranging from 0.724 to 0.942, outperforming the best-performing baseline in each scenario by an average absolute margin of approximately 0.13. WWM also demonstrates improved generalization in un-seen scenarios. In the velocity-generalization setting, WWM achieves an SGCS of 0.673, compared with 0.527 for WiFo and 0.522 for LSTM. In the completely unseen Amsterdam environment, labeled as City Generalization, WWM maintains an SGCS of 0.841, compared with 0.742 for WiFo and 0.586 for LSTM, corresponding to relative improvements of approximately 13.3\% and 43.5\%, respectively. With the aid of multi-modal information, forecasting in the latent space rather than directly in the raw channel domain enables more robust channel prediction across both seen and unseen propagation conditions. A detailed quantitative comparison across models and test scenarios is provided in Table~\ref{tab:csi_prediction_comparison}.

\textbf{CSI compression and feedback}: To evaluate the channel compression capability, we construct a series of experiments, and the results over compression ratios ranging from 1/1024 to 1/128 show that the WWM-based approach achieves the best overall performance compared with QCR-NET~\cite{zhang2023quantization} and CR-NET~\cite{lu2020multi}, as summarized in Table~\ref{tab:full_csi_compression}. Across the in-pattern urban environments (CBD, Etoile, Forbidden City, and Munich), WWM achieves SGCS scores ranging from 0.464 to 0.920, outperforming the most competitive baseline (QCR
NET) by an average absolute SGCS gain of 0.146 across all evaluated compression ratios. Crucially, WWMexhibits robust generalization capabilities under shifting conditions: In the completely unseen city-generalization scenario, WWM achieves an average SGCS of 0.585 across the four compression ratios, compared with 0.348 for QCR-NET and 0.188 for CR-NET, corresponding to relative improvements of approximately 68\% and 212\%, respectively. Even under the extreme compression ratio of 1/1024, WWM attains an SGCS of 0.465 in the unseen city, substantially exceeding QCR-NET (0.249) and CR-NET (0.056). For velocity generalization, WWM also improves over QCR-NET by approximately 9\% on average across the evaluated compression ratios. These results demonstrate that representations learned by WWM can support highly compressed CSI feedback while retaining strong reconstruction accuracy and generalization under unseen propagation environments and mobility conditions, highlighting its potential for practical massive MIMO deployments.

\begin{table}[!h]
    \setlength{\tabcolsep}{4pt}
    \centering
    \small
    \caption{Comprehensive comparison of CSI compression and feedback performance across varying compression ratios and scenarios}
    \begin{tabular}{llcccc}
        \toprule
        \textbf{Model} & \textbf{Scenario} & \textbf{1/1024} & \textbf{1/512} & \textbf{1/256} & \textbf{1/128} \\
        \midrule
        \textbf{WWM} & CBD & \textbf{0.4643} & \textbf{0.5072} & \textbf{0.5033} & \textbf{0.6287} \\
        & Etoile & \textbf{0.6882} & \textbf{0.6840} & \textbf{0.7155} & \textbf{0.9195} \\
        & Forbidden City & \textbf{0.6770} & \textbf{0.6897} & \textbf{0.7222} & \textbf{0.8753} \\
        & Munich & \textbf{0.6311} & \textbf{0.6256} & \textbf{0.6734} & \textbf{0.9104} \\
        & Velocity Gen. & \textbf{0.4611} & \textbf{0.5022} & \textbf{0.5004} & \textbf{0.6087} \\
        & City Gen. & \textbf{0.4650} & \textbf{0.4882} & \textbf{0.5473} & \textbf{0.8403} \\
        \midrule
        \textbf{QCR-NET} & CBD & 0.4435 & 0.4398 & 0.4978 & 0.4916 \\
        & Etoile & 0.4799 & 0.6347 & 0.7187 & 0.7979 \\
        & Forbidden City & 0.2218 & 0.2874 & 0.3647 & 0.4128 \\
        & Munich & 0.5419 & 0.6687 & 0.7557 & 0.8283 \\
        & Velocity Gen. & 0.4515 & 0.4461 & 0.5065 & 0.4973 \\
        & City Gen.  & 0.2493 & 0.3116 & 0.3803 & 0.4499 \\
        \midrule
        \textbf{CR-NET} & CBD & 0.1963 & 0.3206 & 0.4186 & 0.4285 \\
        & Etoile & 0.1374 & 0.2777 & 0.4291 & 0.6087 \\
        & Forbidden City & 0.0679 & 0.1624 & 0.2436 & 0.2778 \\
        & Munich & 0.1805 & 0.3325 & 0.5042 & 0.6551 \\
        & Velocity Gen. & 0.1962 & 0.3291 & 0.4223 & 0.4352 \\
        & City Gen. & 0.0555 & 0.1393 & 0.2609 & 0.2949 \\
        \bottomrule
    \end{tabular}
    \vspace{4pt}
    
    \begin{minipage}{0.78\textwidth}
        \raggedright
        \footnotesize
        A Metric of SGCS is used.
    \end{minipage}
    \label{tab:full_csi_compression}
\end{table}

\begin{table*}[!h]
    \centering
    \caption{beam prediction performance of WWM compared to other baselines across in-distribution and generalization scenarios}
     \begin{tabular}{lcc|cc|cc}
        \hline
        \multirow{2}{*}{Scenario} &
        \multicolumn{2}{c|}{WWM} &
        \multicolumn{2}{c|}{LWM} & \multicolumn{2}{c}{SPMM}\\
        & Top-1 Acc & BG ratio & Top-1 Acc & BG ratio  & Top-1 Acc & BG ratio  \\
        \hline
        CBD & \textbf{0.983} & \textbf{1.000} & 0.927& 0.995& 0.902&0.987\\
        Etoile & \textbf{0.917} & \textbf{0.980} & 0.836& 0.944& 0.778&0.931\\
        Forbidden City & \textbf{0.906} & \textbf{0.981}& 0.854& 0.970 & 0.828&0.963\\
        Munich & \textbf{0.915}& \textbf{0.984}& 0.836& 0.956& 0.767&0.926\\
        Velocity Gen. & \textbf{0.978}& \textbf{0.999}& 0.926& 0.995& 0.857&0.972\\
        
        \hline
    \end{tabular}

    \vspace{4pt}
    \begin{minipage}{0.78\textwidth}
        \raggedright
        \footnotesize
        The Metrics of Top-1 accuracy of predicted DFT codeword and their respective beam gain ratio are used 
    \end{minipage}

    \label{tab:beam_prediction_comparison}
\end{table*}

\textbf{Beam prediction}: Since beam indices are site-dependent labels tied to the BS deployment and antenna configuration, unseen-city evaluation would introduce an additional beam-label alignment problem; we therefore focus on evaluating in in-pattern environments. We evaluate beam prediction using top-1 classification accuracy and beam gain ratio, where beam gain ratio is the channel gain achieved by predicted beam divided by that of theoretical best beam. We compare WWM against two baselines: (i) LWM~\cite{lwm}, a SOTA uni-modal WFM and (ii) SPMM~\cite{alrabeiah2020deep}, a SOTA task-specific. As shown in Table \ref{tab:beam_prediction_comparison}, WWM achieves an average Top-1 accuracy of 94.0\%—a relative increase of 7.3\% over LWM (87.6\%) and 13.6\% over SPMM (82.6\%), demonstrating stronger feature extraction capability. It also demonstrates a very strong  beam gain ratio (99\%), suggesting the predicted beam can always achieve near optimal channel gain.

\begin{figure}[!h]
    \centering
    \includegraphics[width=\columnwidth]{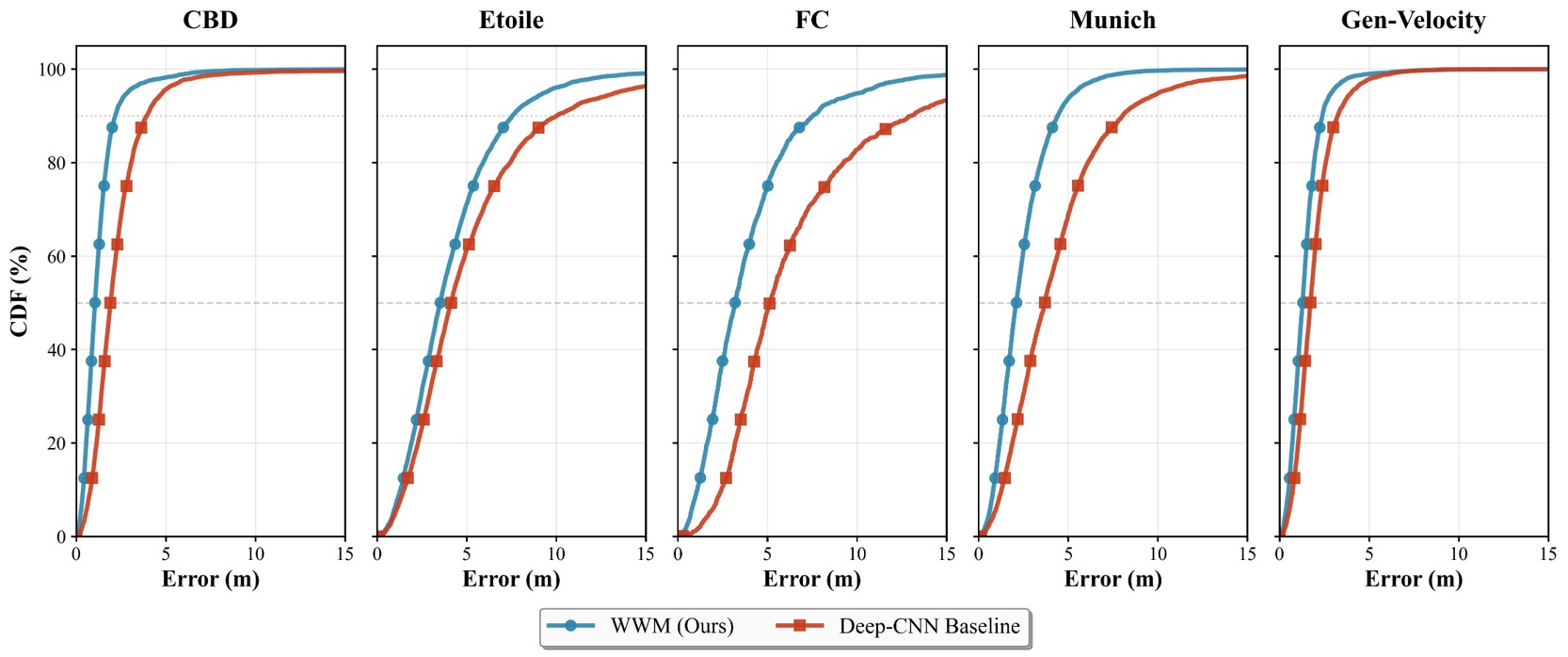}
    \caption{User Localization Performance Comparison}
    \label{fig:localization_results}
\end{figure}

\textbf{User localization}: We quantify localization performance using the cumulative distribution of 2D absolute error in meters, as shown in Fig.~\ref{fig:localization_results}. WWM achieved an average localization error of 2.2 meters, compared with a conventional CNN-based regression baseline model which scored at 4.1 meters, WWM shows a reduction of 46\% in terms of average localization error across all tested scenarios, highlighting its ability to extract position-relevant features directly from raw CSI and 3D point clouds. 

\textbf{Ablation Study:} To investigate the effectiveness of the proposed MMoE design, we conducted ablation experiment to evaluate the performance of WWM without MMoE design (combining 3 Modality-specific feed-forward experts into one) across different scenarios by analyzing the SGCS of CSI prediction at two future time steps. As shown in Table \ref{tab:ablation_multi-modal}, WWM achieves consistently higher SGCS values than the one without MMoE across both in-distribution and generalization scenarios (outperforming by an average SGCS of 0.067, from 0.765 to 0.832), demonstrating the robustness of the modality-specific expert design under diverse propagation conditions.

To further quantify the contribution of multi-modal fusion to the WWM, we conducted ablation experiment in which the model was pre-trained using only CSI data, with 3D point cloud and user trajectory modalities masked. This uni-modal variant preserves the same JEPA pre-training scheme and model architecture (despite removing 2 un-used modality-specific expert FFNs), differing only in the absence of auxiliary modalities. When evaluated on the CSI temporal prediction task, as shown in Table \ref{tab:ablation_multi-modal}, the uni-modal model exhibited consistently lower prediction SGCS (a decrease from 0.765 to 0.740 on average across all scenarios). These results indicate that geometric context and motion cues provide essential complementary information that cannot be recovered from CSI alone.

        
    

\begin{table*}[!h]
    \centering
    \caption{Ablation study comparing the full multi-modal WWM with variants without MMoE and without auxiliary modalities.}
    \begin{tabular}{lccc|ccc|ccc}
        \hline
        \multirow{2}{*}{Scenario} &
        \multicolumn{3}{c|}{WWM} &
        \multicolumn{3}{c|}{WWM (w/o MMoE)} &
        \multicolumn{3}{c}{WWM-Unimodal} \\
        & T=15 & T=16 & Avg. 
        & T=15 & T=16 & Avg. 
        & T=15 & T=16 & Avg. \\
        \hline
        CBD 
        & 0.707 & 0.741 & \textbf{0.724} 
        & 0.580 & 0.562 & 0.571
        & 0.541 & 0.535 & 0.538 \\
        
        Etoile 
        & 0.931 & 0.919 & \textbf{0.925} 
        & 0.917 & 0.902 & 0.909
        & 0.888 & 0.868 & 0.878 \\
        
        Forbidden City 
        & 0.943 & 0.941 & \textbf{0.942} 
        & 0.890 & 0.880 & 0.885
        & 0.882 & 0.864 & 0.873 \\
        
        Munich 
        & 0.896 & 0.881 & \textbf{0.888} 
        & 0.877 & 0.857 & 0.867
        & 0.831 & 0.798 & 0.815 \\
        
        Velocity Gen. 
        & 0.659 & 0.688 & \textbf{0.673} 
        & 0.521 & 0.510 & 0.515
        & 0.495 & 0.502 & 0.498 \\
        
        City Gen. 
        & 0.854 & 0.829 & \textbf{0.842} 
        & 0.853 & 0.827 & 0.840
        & 0.851 & 0.825 & 0.838 \\
        \hline
    \end{tabular}
        
    \vspace{4pt}
    \begin{minipage}{0.78\textwidth}
        \raggedright
        \footnotesize
        A Metric of SGCS is used.
    \end{minipage}
    
    \label{tab:ablation_multi-modal}
\end{table*}

To validate the deployment feasibility of WWM in RAN hardware, we quantified the end-to-end inference latency of WWM. For the frozen WWM backbone (including encoder and predictor) paired with decoder heads, the average single-sample end-to-end inference latency is 8.5 ms on an NVIDIA RTX 4090 24GB GPU under mixed-precision bfloat16 inference. 

\subsection{WWM's generalization capability in Real-world measurement}
As the fifth task, predicting CSI in frequency-domain serves as a critical test of model transferability from large-scale simulated CSI pre-training to real-world measurements. Specifically, WWM is first pre-trained on simulated CSI data, and then fine-tuned using a comparatively small real-world measured CSI dataset collected from the 6G prototype system. Given CSI measured on four SRS Physical Resource Blocks (PRBs) in frequency domain, the model predicts the CSI on adjacent four PRBs within the same time window. 

Despite the substantial gap between simulated and measured channel—arising from hardware impairments, measurement noise, and mismatched feature statistics—WWM rapidly adapts to real environments and outperforms the baseline WFM method WiFo\cite{wifo1} and task-specific model C-Mixer~\cite{chen2024channel}. As shown in Table ~\ref{tab:srs_prediction_comparison}, it achieves improvements of 15.9\% and 27.6\% in NMSE, and 3.1\% and 4.0\% in SGCS, respectively.
These results indicate that the model learns transferable representations of wireless dynamics that extend beyond the specific characteristics of ray tracing simulations, and that limited real measurements (13.2\% of simulated pre-training data volume) are sufficient to specialize the model to real-world systems. This observation suggests a practical pathway for deploying foundation models in future 6G networks, where large-scale simulation can be leveraged for pre-training, and lightweight fine-tuning on sparse real data enables fast adaptation to new environments and system configurations.

\begin{table}[!h]
    \centering
    \caption{CSI frequency-domain prediction performance of WWM compared to other baselines}
     \begin{tabular}{lcc|cc|cc}
        \hline
        \multirow{2}{*}{Scenario} &
        \multicolumn{2}{c|}{WWM} &
        \multicolumn{2}{c|}{WiFo} & \multicolumn{2}{c}{C-Mixer}\\
        & NMSE & SGCS & NMSE & SGCS  & NMSE &SGCS  \\
        \hline
        PRB4 & \textbf{0.212} & \textbf{0.932} & 0.252& 0.922& 0.296 &0.918  \\
        PRB5 & \textbf{0.218} & \textbf{0.918} & 0.261& 0.895& 0.308 &0.892  \\
        PRB6 & \textbf{0.260} & \textbf{0.914} & 0.309& 0.882& 0.349 &0.875  \\
        PRB7 & \textbf{0.243} & \textbf{0.902} & 0.287& 0.858& 0.334 &0.843  \\
        Avg. & \textbf{0.233} & \textbf{0.917} & 0.277& 0.889& 0.322 &0.882  \\
        
        \hline
    \end{tabular}
    \label{tab:srs_prediction_comparison}
\end{table}

\section{Conclusion}

This paper presented a WWM for AI-native 6G networks, aiming to move wireless AI from fragmented task-specific models toward a unified foundation-model paradigm. By integrating CSI, 3D point clouds, and user trajectories through a multi-modal mixture-of-experts architecture, WWM learns a shared latent representation that captures both electromagnetic propagation and environmental dynamics. To support this design, we constructed a large-scale hybrid dataset with more than 800 thousand samples, combining ray-tracing simulations and real-world 6G prototype measurements.

Extensive experiments demonstrate that WWM achieves strong performance across channel prediction, beam management, and localization tasks, with clear gains in cross-site generalization. The results further show that multi-modal pre-training provides essential geometric and mobility priors, enabling the model to better capture complex channel evolution than uni-modal or task-specific baselines. Although the initial pre-training introduces additional cost, it can be amortized across multiple sites and downstream tasks, reducing the long-term burden of deploying and maintaining isolated models. Future work will focus on reducing reliance on real-time 3D geometry, extending WWM to massive multi-user scenarios, and improving efficiency through model compression and lightweight inference.

\section*{Acknowledgment}

This work was supported by the National Science \& Technology Major Project of China under Project GXB-2-2025-5. The authors thank Yu Wang, Xu Xue, Yiwei Yan, Yang Wang, Baichuan Zhao, Zihe Wang, Tiancai Wei, Shihao Zhang, Xiaoyan Chen, and Hongjun He for technical validation and field-measurement support.

\section*{Data and model availability}\label{sec5}

Once the paper is accepted, all datasets and model checkpoints used in this study will be publicly available via \href{https://zenodo.org/communities/wwm/}{https://zenodo.org/communities/wwm/}. Below are private link for review.

\textbf{Datasets.} 
The multi-modal wireless datasets include both simulated and real-world measurements:

(i) a ray-traced simulated dataset for large-scale pre-training (\href{https://zenodo.org/records/18785355?token=eyJhbGciOiJIUzUxMiJ9.eyJpZCI6IjZhMzc3NTM2LWI4M2YtNGM5MS04YWEzLTA1ZGFjNjIyNTVlZiIsImRhdGEiOnt9LCJyYW5kb20iOiJiNmRiMDUxNzA3YmVlNDRiODlhZTg2YzA3NzUyODNjYyJ9.3VGCN51icU-FPupYE4F2hjZpJWvLmdcZmJPGgGtQX2cBRIwPOlVNGW45zgUUD3pwZblLob9YJ9S54Xj19JQr8Q
}{https://doi.org/10.5281/zenodo.18785355}); 

(ii) a simulated dataset for evaluating the pre-trained model (\href{https://zenodo.org/records/22023746?token=eyJhbGciOiJIUzUxMiJ9.eyJpZCI6ImY5NmJhMWFjLTc3ZWEtNGFlNi05MDYyLTc0YTFjNTA2NDc5NCIsImRhdGEiOnt9LCJyYW5kb20iOiI1ZjIxZTI1OGVlYjU1ZmUxOGZkNWEyMGE0MTQ2ZWU3YyJ9.MaLwePJCoynxRMf44Hq5tFnaMkYThNSSsO_91DerHoWkhfSPve37iSJplIxJLnUQyooYnmKm41aRW16G38YswQ
}{https://doi.org/10.5281/zenodo.22023746}); 

(iii) three-dimensional point cloud data describing the environments of all scenarios (\href{https://zenodo.org/records/22023546?token=eyJhbGciOiJIUzUxMiJ9.eyJpZCI6ImIzNTIzN2FkLWU5ZjAtNDkyOC1hNmY3LTIzZmE0NWIxZGNlMyIsImRhdGEiOnt9LCJyYW5kb20iOiJiYWUwOTM0ZGFmZDVmNjEzNjIxYTFkYTNkYmUxNTZiNSJ9.Vpc1-ZUjWFxodxc_-h02MXXfQDZOGVepAkvxGF4FS0-IUgoWoMdyYKO_54JaS2TD9sYIAjQDZs6FSQfamkyfGQ
}{https://doi.org/10.5281/zenodo.22023546}); 

(iv) a ray-traced dataset at 2.6\,GHz and 6.6\,GHz for beam prediction tasks (\href{https://zenodo.org/records/18797221?token=eyJhbGciOiJIUzUxMiJ9.eyJpZCI6ImQ4NTU4NmM1LWQ0OTItNDUyOC1hMGYxLTZjNTcxODc2ZDM5ZiIsImRhdGEiOnt9LCJyYW5kb20iOiJkYzAzNGQ0YmEyMDRkNTM3ODM5MWFlYzhhMWRmNzk2YyJ9.31M4eKNNy4JjC_uwnw66JbvjpajpiYWWqP3dEzAiXmzw4megmnScwga-w4iwvM6o1xlXgOcdYE_jdApNteAGfw}{https://doi.org/10.5281/zenodo.18797221}); and 

(v) a real-world field measurement dataset based on uplink CSI collected from a 6G prototype system using sounding reference signals (SRS) \href{https://zenodo.org/records/18783399?token=eyJhbGciOiJIUzUxMiJ9.eyJpZCI6Ijc2OWE1OTE5LWMyYTMtNGE5OS1iM2U5LTA4ODEwYjQyYWEyZiIsImRhdGEiOnt9LCJyYW5kb20iOiJiNzA5MTcxZGM0Y2NlOTk5ZTJmMzcwOTZiYzJlMjYzOSJ9.RZ8u5Xiu1ce784S5d-LSN4_QBgE4DK3wn88rdUa4A8Nd2Jw8uo2KNp5eSBhrxXL4O1ZhUIkSkuZXN6llS45OoA
}{https://doi.org/10.5281/zenodo.18783399}).

\textbf{Models.} 
The pre-trained Wireless World Model (WWM) and downstream task checkpoints will be available after accept at {https://doi.org/10.5281/zenodo.22299088}.    

\section*{Code availability}\label{sec6}
Once the paper is accepted, the model pre-training, downstream task training and testing code will be publicly available via GitHub at {https://github.com/Wireless-World-Model/WWM-V1}.  

\bibliographystyle{IEEEtran}
\bibliography{JSAC_wwm_bib}

\end{document}